\documentclass{aastex63}

\newcommand{\cmnt}[1]{}

\usepackage{amsmath}
\usepackage{amssymb}
\usepackage{tabularx}

\graphicspath{{./}{figures/}}

\begin{document}

\title{Effect of dust in circumgalactic haloes on the cosmic shear power spectrum}

\author[0000-0002-1717-4424]{Makana Silva}
\affiliation{The Center for Cosmology and AstroParticle Physics (CCAPP) \\
Physics Research Building
191 West Woodruff Avenue \\
Columbus, OH 43210, USA}

\author[0000-0002-2951-4932]{Christopher Hirata}
\affiliation{The Center for Cosmology and Astroparticle Physics (CCAPP) \\
Physics Research Building
191 West Woodruff Avenue \\
Columbus, OH 43210, USA}

\begin{abstract}
Weak gravitational lensing is a powerful statistical tool for probing the growth of cosmic structure and measuring cosmological parameters. However, as shown by studies such as \cite{2010MNRAS.405.1025M}, dust in the circumgalactic region of haloes dims and reddens background sources. In a weak lensing analysis, this selects against sources behind overdense regions; since there is more structure in overdense regions, we will underestimate the amplitude of density perturbations $\sigma_8$ if we do not correct for the effects of circumgalactic dust. To model the dust distribution we employ the \textit{halo model}. Assuming a fiducial dust mass profile based on measurements from \cite{2010MNRAS.405.1025M}, we compute the ratio $Z$ of the systematic error to the statistical error for a survey similar to the {\slshape Nancy Grace Roman Space Telescope} reference survey (2000 deg$^2$ area, single-filter effective source density 30 galaxies arcmin$^{-2}$). For a waveband centered at $1580$ nm ($H$-band), we find that $Z_{H} = 0.37$. For a similar survey with waveband centered at $620$ nm ($r$-band), we also computed $Z_{r} = 2.8$. Within our fiducial dust model, since $Z_{r} > 1$, the systematic effect of dust will be significant on weak lensing image surveys. We also computed the dust bias on the amplitude of the power spectrum, $\sigma_{8}$, and found it to be for each waveband $\Delta \sigma_8/\sigma_8 = -3.1\times 10^{-4}$ ($H$ band) or $-2.2\times 10^{-3}$ ($r$ band) if all other parameters are held fixed (the forecast {\slshape Roman} statistical-only error $\sigma(\sigma_8)/\sigma_8$ is $9\times 10^{-4}$).  
\end{abstract}


\section{Introduction} \label{sec:intro}

One of the main goals in modern cosmology is to characterize and understand the accelerated expansion of the Universe \citep{1998AJ....116.1009R,1999ApJ...517..565P}. A key question driving the current generation of experiments is whether the mechanism responsible for this expansion is a cosmological constant, a new dynamical field, or a modification to general relativity \citep[e.g.,][]{2006astro.ph..9591A,NAP12951,2013PhR...530...87W}. One promising method of studying cosmic acceleration is with weak gravitational lensing -- the subtle change in shape of background galaxies caused by perturbations along the line of sight (see \citealt{2018ARA&A..56..393M} for a recent review).
Because lensing is sensitive to the gravitational potential perturbations in the Universe, it complements other techniques such as supernovae and baryon-acoustic oscillations that tightly constrain the background geometry. One can probe the growth of structure as a function of cosmic time by measuring the weak lensing shear correlation function (often complemented by galaxy-shear correlations and galaxy clustering observables) as a function of scale and source redshift. Modified gravity theories can then be constrained by using initial conditions anchored to cosmic microwave background observations and predicting the amplitude of structure (see \citealt{2019PhRvD..99l3505A, 2019PhRvD..99h3512F, 2021arXiv210414515L} for recent analyses).

Because the weak lensing shear signal is small ($\sim 1\%$; \citealt{2003moco.book.....D}), we require large samples of source galaxies \citep{2001PhRvD..65b3003H} and tight control of systematic errors to measure it. Even larger samples are required in order to test models of the growth of structure; for example, even the large change from the Einstein-de Sitter ($\Omega_m=1$) to the presently favored $\Lambda$CDM ($\Omega_m=0.3$) model only reduces the growth function by 20\% at $z=0$ \citep[e.g.][]{1992ARA&A..30..499C}. The initial detections of weak lensing were carried out with thousands to hundreds of thousands of source galaxies \citep{2000MNRAS.318..625B, 2000A&A...358...30V, 2000Natur.405..143W, 2001ApJ...552L..85R}. Both the sample size and effort devoted to systematics mitigation expanded with the ``Stage II'' Canada-France-Hawaii Telescope Lensing Survey \citep{2013MNRAS.430.2200K, 2013MNRAS.432.2433H}. The current ``Stage III'' surveys, reaching few percent precision on the amplitude $S_8$, include the Dark Energy Survey (DES, \citealt{2018PhRvD..98d3528T, 2021arXiv210513543A,2021arXiv210513544S}), the Hyper Suprime Cam (HSC, \citealt{2019PASJ...71...43H, 2020PASJ...72...16H}), and the KiloDegree Survey (KiDS, \citealt{2020A&A...633A..69H, 2020A&A...634A.127A}).
The upcoming generation of ``Stage IV'' surveys -- \textit{Euclid} \citep{2011arXiv1110.3193L},  the Vera Rubin Observatory Legacy Survey of Space and Time (LSST, \citealt{2012arXiv1211.0310L}), and the \textit{Nancy Grace Roman Space Telescope} \citep{2019arXiv190205569A}
-- will aim for precision of a fraction of a percent, enabling robust measurements of the amplitude of gravitational potential perturbations across a range of scales and redshifts.

Dust along the line of sight can reduce the apparent brightness of a background source and thus affect whether it passes the magnitude or signal-to-noise cuts for inclusion in a lensing sample. Even apparently ``empty'' lines of sight pass through galactic haloes and could contain dust, and some of the early studies of galaxy-quasar correlations considered dust extinction as a possible systematic \citep[e.g.][]{1997AJ....114.1728F, 2005ApJ...633..589S}. \cite{2010MNRAS.405.1025M} used the well-calibrated photometry from the Sloan Digital Sky Survey, covering $u$ through $z$ bands, to fit both the magnification and dust reddening contributions to the correlation function of $z\sim 0.3$ galaxies and $z\gtrsim 1$ quasars. They find a power law-like dust signal extending from the inner halo ($r_p \sim 20$ kpc) out to the large-scale clustering regime (several Mpc). This excess of diffuse dust in high density regions implies that we will observe fewer source galaxies behind a higher-density region. Since matter density perturbations grow faster in higher-density regions, this selection effect will lead to an underestimate of the weak lensing power spectrum and of the amplitude of perturbations $\sigma_{8}$. Here we present a preliminary calculation of the impact of the dust systematic on the weak lensing power spectrum and on the determination of the amplitude. For this analysis, we do not consider the interaction of dust corrections with other corrections such as multiple deflections \citep{2002ApJ...574...19C} and reduced shear \citep{2009ApJ...696..775S}, since these would depend on higher order moments of the density field than the 3rd order moments that dominate this work.\footnote{The corrections with dust interacting with reduced shear and multiple deflections would be of at least 4th order in the density field. For the mathematically similar integrals of the reduced shear interacting with the multiple deflection correction, \citet{2010A&A...523A..28K} found that the 4th order contributions to the observed shear power spectrum are small compared to the 3rd order terms. We leave for future work the task of checking whether the same is true for dust.}

This paper is outlined as follows: in Sec. \ref{sec:theory} we set up the basic formulation for weak lensing, how to compute the lowest order correction of interest to the shear power spectrum, and the theoretical basis for describing the dust: the halo model approach. In Sec. \ref{sec:dust_measurement} we describe how we implement measured dust masses in halos into our model. In Sec. \ref{sec:results} displays the results of our calculations. In Sec. \ref{sec:dicussion} we analysis our results and discuss the implication of the estimated error due to dust. Appendix~\ref{sec: Appendix_A} contains some formal discussion on error metrics and propagation.

We used the cosmological parameters in Table 1 in \cite{2020A&A...641A...6P}.

\section{Theory \& Background} \label{sec:theory}
\subsection{Weak Lensing Formalism}
In order to study the effects of dust on the selection of weakly lensed galaxies, we need to calculate the corrections to the shear power spectrum. Before we do this, we develop the relevant formalism in weak lensing that we will use in our analysis.

\begin{center}
\begin{figure}[h!]
\includegraphics[scale=0.6]{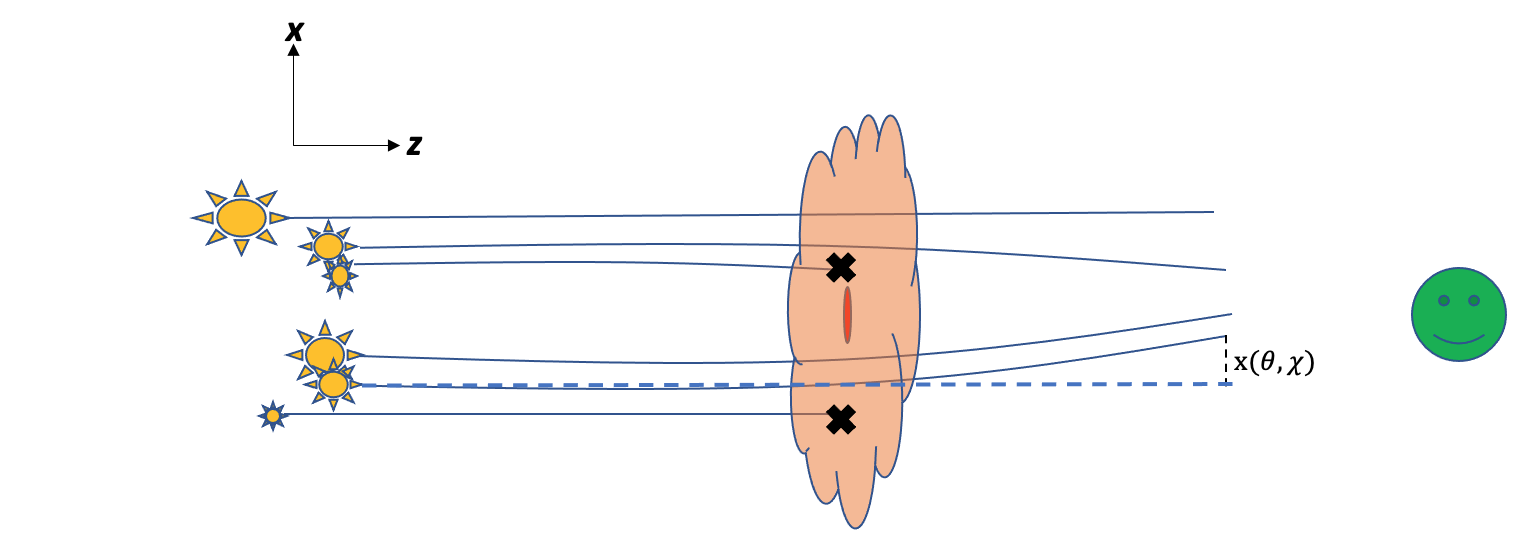}
\caption{The light rays from the background sources (yellow stars) are being deflected  by the gravitational field of the foreground matter (red lens).This creates a distorted image in the observer frame (green smiley face). The deformation is quantified by $\vec{x}(\theta,\chi)$, which measures how much the distorted light rays separate from the undeflected rays (dashed lines). The dust (orange cloud) within the galactic halo would affect the incoming light from the background sources by attenuating the light. This will result in some of the background light sources going undetected (black X's).
\label{fig:Lensing}}
\end{figure}
\end{center}

Weak lensing is the result of background light being distorted due to foreground gravitational fields, as depicted in Figure~\ref{fig:Lensing}.

Weak lensing is mathematically described using two quantites, shear and convergence. We can think of a lensed image as being a result of a transformation from the true image to the distorted (lensed) image. We can establish a connection between the true image (unlensed) and the observed image (lensed) by calculating the deformation matrix defined in \cite{1998MNRAS.296..873S}:
\begin{align}
\label{deformation_matrix1}
    \vec{x}(\vec{\theta},\chi) &= d_{A}(\chi)\vec{\theta} - \frac{2}{c^2}\int_{0}^{\chi}d\chi'd_{A}(\chi - \chi')\nabla_{\bot}\Phi(\vec{x}(\vec{\theta},\chi'),\chi'),
\end{align}
where $\vec{x}(\vec{\theta},\chi)$ is the transverse location of the source without lensing effects, $c$ is the speed of light, $\vec{\theta}$ is the observed angular position in the sky, $d_{A}(\chi)$ is the angular diameter distance, $\chi$ is the comoving radial distance, and $\nabla_{\bot}\Phi(\vec{x}(\vec{\theta},\chi'),\chi')$ is the derivative of the Newtonian gravitational potential of the lens in the transverse (sky) coordinates (in comoving Mpc). For most weak lensing calculations, since the strength of the deformation is relatively weak, and we work on a spatially flat background, we will set $d_{A}(\chi) = \chi$.

We can now define the deformation matrix by computing the Jacobian, the derivative of the transverse position of the source (``true'' position) with respect to the observed image position, $A_{ij} = \partial \beta_{i}/ \partial \theta_{j}$, where $\vec{\beta} \equiv \vec{x}/d_{A}(\chi)$ and the lower Latin indices label sky coordinates.

For weak lensing, we can expand $A_{ij}(\vec{\theta},\chi)$ up to first order in $\Phi(\vec{x}(\vec{\theta},\chi'))$ and obtain
\begin{align}
\label{deformation_matrix3}
    A_{ij}(\vec{\theta},\chi) &= \delta_{ij} - \frac{2}{c^2}\int_{0}^{\chi}d\chi' \frac{(\chi - \chi')\chi'}{\chi} \frac1{\chi'{^2}} \partial_{i}\partial_{j}\Phi(\vec{x}(\vec{\theta},\chi'),\chi'),\\
    &= \delta_{ij} - \psi_{,ij},\\
    \label{eq: lensing_potential}
    \psi &\equiv \frac{2}{c^2}\int_{0}^{\chi}d\chi' \frac{\chi - \chi'}{\chi\chi'} \Phi(\vec{x}(\vec{\theta},\chi'),\chi') = \frac{2}{c^2}\int_{0}^{\chi}d\chi'\, W(\chi',\chi) \Phi(\vec{x}(\vec{\theta},\chi'),\chi'),
\end{align}
where Eq. (\ref{eq: lensing_potential}) defines the lensing potential and the subscripts indicate derivatives in angular sky coordinates (in radians). We have defined the window function $W(\chi',\chi) =\left(1/\chi' - 1/\chi \right)\Theta(\chi - \chi')$, where $\Theta$ is the Heaviside step function. We can now define the deformation matrix from the observed (lensed image) to the true image (unlensed image),
\begin{align}
\label{lensing_transform}
    \begin{pmatrix}
    \theta_{t,1}\\
    \theta_{t,2}
    \end{pmatrix} &= \begin{pmatrix}
    1 - \gamma_{1} -\kappa & -\gamma_{2}\\
    -\gamma_{2} & 1 + \gamma_{1} - \kappa
       \end{pmatrix} \begin{pmatrix}
    \theta_{o,1}\\
    \theta_{o,2}
    \end{pmatrix},\\
    \label{eq: kappa}
    \kappa &= \frac{1}{2}(\psi_{,11} + \psi_{,22}),\\
    \label{eq:gamma_1}
    \gamma_{1} &= \frac{1}{2}(\psi_{,11} - \psi_{,22}),\\
    \label{eq: gamma_2}
    \gamma_{2} &= \psi_{,12},
\end{align}
where $\{\theta_{o,1},\theta_{o,2}\}$ are the observed angular coordinates of the object in the sky, $\{\theta_{t,1},\theta_{t,2}\}$ are the true coordinates in the sky, $\kappa$ is the convergence and $\gamma_{1,2}$ represent the Cartesian components of shear. Physically, shear represents the change in ellipticity. Convergence tells you how magnified the image has become after passing through the lens. For weak lensing, $\kappa \ll 1$.

From here on, we shall take $c=1$ and we will use the standard convention for 2D angular Fourier transforms in the flat-sky approximation:
\begin{equation}
\label{Fourier}
    f(\vec{\theta}) = \int \frac{d^{2}\vec{l}}{(2\pi)^2}\widetilde{f}(\vec{l})e^{i\vec{l}\cdot\vec{\theta}}
    ~~~\leftrightarrow~~
    \widetilde{f}(\vec{l})=\int d^{2}\vec{\theta}\,{f}(\vec{\theta})e^{-i\vec{l}\cdot\vec{\theta}}
    ,
\end{equation} 
and define the convolution of two Fourier-domain functions by
\begin{equation}
\label{convolution}
    [\widetilde{f} * \widetilde{g}](\vec{l}) = \int \frac{d^{2}\vec{l'}}{(2\pi)^2}\widetilde{f}(\vec{l})\widetilde{g}(\vec{l} - \vec{l'}),
\end{equation} 
where $*$ denotes the convolution of two functions and tilde represents the Fourier transformed function. We will also follow the convention that Greek letters are to index redshift slices and capital Roman letters denote shear components.

Now that we have defined the shear and convergence, we can describe how they are relevant to computing corrections to the shear power spectrum. 

\subsection{Computing the Correction to the Shear Power Spectrum}
\label{Third_order_correction}

Since we can only observe the shear field at the positions of source galaxies, the shear is statistically weighted by the number of galaxies we observe. In our case, cosmic dust correlates the source density with the lensing signal itself: sources behind large-scale overdensities suffer more extinction and are less likely to be selected. Since there is more small-scale structure in large-scale overdensities (due to non-linear evolution), we expect this selection effect to lower our estimate of the lensing amplitude, and produce a downward bias in $\sigma_8$ (with all other parameters fixed).

These selection effects can be treated by defining a weighted shear $\gamma_{I,\rm weight} = (1+\Delta N/N) \gamma_I$, where $\Delta N/N$ is the perturbation to the source density \citep{2009PhRvL.103e1301S}.\footnote{The ``weighting'' effect is the lowest-order effect of a variation in selection function in the survey. A detailed discussion of the subtleties in shear correlation functions with varying source number density can be found in \citet{2009ApJ...702..593S}.} In our case, this gives the expression for the first order corrections to the shear, $\gamma^{(1)}$, by
\begin{align}
\label{eq: wt_reduced_shear_real}
    \gamma^{(1)}_{I,weight}(\vec{\theta},z_{\alpha}) = \left(1+\frac{d\ln N}{d\ln F}(z_{\alpha})\Delta \tau(\vec{\theta},z_{\alpha}) \right)\gamma^{(1)}_{I}(\vec{\theta},z_{\alpha}),
\end{align}
where $\frac{d\ln N}{d\ln F}(z_{\alpha})$ is the number of sources detected with measured flux, $F$, at redshift $z_{\alpha}$, and $\Delta \tau$ is the dust optical depth perturbation, which will be specified later. Since we are interested in correction to the shear power spectrum, we will take the Fourier transform of Eq. (\ref{eq: wt_reduced_shear_real}) to get
\begin{align}
\label{eq:wt_reduced_shear_Fourier}
    \widetilde{\gamma}^{(1)}_{I,weight}(\vec{l},z_{\alpha}) = \widetilde{\gamma}^{(1)}_{I}(\vec{l},z_{\alpha})+\frac{d\ln N}{d\ln F}(z_{\alpha})[\widetilde{\Delta \tau} *\widetilde{\gamma}^{(1)}_{I}](\vec{l},z_{\alpha}).
\end{align}
 The term $\frac{d\ln N}{d\ln F}(z_{\alpha})$ appears since dust reduces the brightness of the background sources thereby reducing the number of sources detected at a particular flux. Since the number of observed sources decrease with increasing brightness, we know that $\frac{d\ln N}{d\ln F}(z_{\alpha}) < 0$. We can compute the correlation between the weighted shears at different redshift slices by
\begin{equation}
\label{eq:correlation_def}
    \langle \widetilde{\gamma}^{(1)}_{I,weight}(\vec{l},z_{\alpha}) \widetilde{\gamma}^{(1)}_{I,weight}(\vec{l''},z_{\beta})\rangle = (2\pi)^2 \delta_{D}(\vec{l} + \vec{l''})C_{ll''}^{wt,\alpha \beta},
\end{equation}
where
$C_{ll''}^{wt,\alpha \beta} = C_{ll''}^{\alpha \beta} + \Delta C_{ll''}^{\alpha \beta}$.
If we expand the left hand side of Eq. (\ref{eq:correlation_def}) to the lowest order in dust ($\Delta \tau$), we will find that two terms: (1) the shear power spectrum (\citealt{2010A&A...523A..28K}, Eq. 15) and (2) the lowest order correction defined by
\begin{align}
\label{eq:lowest_correction}
    (2\pi)^2\delta_{D}(\vec{l} + \vec{l''})\Delta C_{ll''}^{\alpha \beta} = \langle \frac{d\ln N}{d\ln F}(z_{\beta}) \widetilde{\gamma}^{(1)}_{I}(\vec{l},z_{\alpha})(\widetilde{\gamma}^{(1)} * \widetilde{\Delta \tau})_{E}(\vec{l''},z_{\beta}) \rangle + \langle \frac{d\ln N}{d\ln F}(z_{\alpha}) \widetilde{\gamma}^{(1)}_{I}(\vec{l''},z_{\beta})(\widetilde{\gamma}^{(1)} * \widetilde{\Delta \tau})_{E}(\vec{l},z_{\alpha}) \rangle. 
\end{align}
We can see on the right hand side of Eq. (\ref{eq:lowest_correction}) that these quantities in the ensemble brackets are third order in the correction, two products of shear and one of dust. We also note that given regime of interest, we will assume the slope $\frac{d\ln N}{d\ln F}(z_{\alpha})$ is a constant over the sky average, which means we can take them out of the average brackets (the spatial variation of $d\ln N/d\ln F$ contains a higher-order correlation function, which would be small anyway because it contains a correlation of a perturbation at the source with a perturbation along the line of sight). 
Equation~(\ref{eq:lowest_correction}) also shows that the lowest correction to the observed shear power spectrum contains a 3-point statistic: the correlation of two linear shears and an optical depth. The bispectrum has appeared in related problems with weighted shear power spectra, such as the reduced shear power spectrum \citep{2009ApJ...696..775S, 2020A&A...636A..95D}.

To continue, we recall the relations between the first-order shear, convergence, and potential:
\begin{equation}
\label{eq: shearEmode}
\begin{split}
    \widetilde{\kappa}^{(1)}(\vec{l},z_s) = -l^{2}\int_{0}^{\chi_{s}}d\chi W(\chi,\chi_{s})\widetilde{\phi}(\vec{l};\chi),
    \\
    \widetilde{\gamma}^{(1)}_{I}(\vec{l},z_s) = T_{I}(\vec{l})\widetilde{\kappa}^{(1)}(\vec{l},z_s),
    \\
    \widetilde{\gamma}^{(1)}_{E}(\vec{l},z_s) = \widetilde{\kappa}^{(1)}(\vec{l},z_s),~~~{\rm and}
    \\
    \widetilde{\gamma}_{E}(\vec{l},z_s) = \delta_{IJ}T_{I}(\vec{l}) \widetilde{\gamma}_{J}(\vec{l},z_s),
\end{split}
\end{equation}
where $\widetilde{\phi}$ is the Fourier transform of the Newtonian potential in angular coordinates, $\vec{l}$ are the angular modes on the sky ($\vec{k}\chi$), $\phi_{l}$ is the azimuthal angle of $\vec{l}$, $\widetilde{\kappa}^{(1)}$ and $\widetilde{\gamma_{I}}^{(1)}$ are the first order approximations of the Fourier transforms of the convergence and Cartesian component of shear, $z$ is redshift of the source, and $\chi$ is the comoving radial distance from observer as defined in \cite{2010A&A...523A..28K}. The first three expressions in Eq. (\ref{eq: shearEmode}) are first order approximations defined by substituting Eq. (\ref{eq: lensing_potential}) in Eq. (\ref{eq: kappa}) and Eq. (\ref{eq:gamma_1}). The fourth expression is the transforming from the Cartesian components of shear to the tangential ($E$) component; the rotation coefficients are
\begin{equation}
\label{eq: transform_Emode}
    T_{1}(\vec{l}) = \cos{2\phi_{l}}
    ~~{\rm and}~~
    T_{2}(\vec{l}) = \sin{2\phi_{l}},
\end{equation}
where the factor of ``2'' is present because the shear is a spin 2 field. Finally, we express the dust profile as
\begin{equation}
\label{eq: dust_profile}
\begin{split}
    \Delta \tau(\vec{l},\chi) = \int^{\chi}_{0} \frac{d\overline{\tau}}{d\chi'}\delta_{d}(\vec{l},\chi')d\chi'.
\end{split}
\end{equation}
The expression for $\Delta \tau$ is a measure of the change in mean optical depth out to the observing point by projecting a three dimensional field ($\delta_{d}(\vec{l},\chi)$, the dust overdensity) into a two dimensional measurement via integrating along the line of sight. The mean optical depth per unit length, $\frac{d\overline{\tau}}{d\chi}$, requires the mean density of dust, which we will define in Sec. \ref{subsec: dust_bispectrum}.

In order to compute the correction of interest, $\Delta C_{ll'}^{\alpha \beta}$, we need to compute the quantity in the angular brackets in Eq. (\ref{eq:lowest_correction}) by substituting Eqs. (\ref{eq: shearEmode}--\ref{eq: dust_profile}) into Eq.~(\ref{eq:lowest_correction}):
\begin{gather}
\label{eq: reducedshear1}
    \langle \widetilde{\gamma}^{(1)}_{E}(\vec{l},z_{\alpha}) (\widetilde{\gamma}^{(1)} * \widetilde{\Delta \tau})_{E}(\vec{l''},z_{\beta})\rangle  = \langle \widetilde{\kappa}^{(1)}(\vec{l},z_{\alpha}) \delta_{LM}T_{L}(\vec{l''})(\widetilde{\gamma}_{M}^{(1)}*\widetilde{\Delta \tau})(\vec{l''},z_{\beta}) \rangle,\\
\label{eq: reducedshear2}
    =\langle \widetilde{\kappa}^{(1)}(\vec{l},z_{\alpha}) \delta_{LM}T_{L}(\vec{l''})\int \,\frac{d^2\vec{l'''}}{(2\pi)^2} T_{M}(\vec{l'''})\widetilde{\kappa}^{(1)}(\vec{l'''},z_{\beta}) \widetilde{\Delta \tau}(\vec{l''} - \vec{l'''})\rangle,\\
\label{eq: reducedshear3}
    = \langle \widetilde{\kappa}^{(1)}(\vec{l},z_{\alpha})\int\frac{d^2\vec{l'''}}{(2\pi)^2} \cos(2\phi_{l''} - 2\phi_{l'''})\widetilde{\kappa}^{(1)}(\vec{l'''},z_{\beta}) \widetilde{\Delta \tau}(\vec{l''} - \vec{l'''}) \rangle .
\end{gather}
We will use the Limber approximation found in \cite{2010A&A...523A..28K} in their Eq. (21) in order to relate the expectation value in Eq. (\ref{eq: reducedshear4}) to the matter-matter-dust bispectrum in Eq. (\ref{eq: reducedshear5})
\begin{gather}
\label{eq: reducedshear4}
     = \int_{0}^{\chi_{\alpha}} \,d\chi \int_{0}^{\chi_{\beta}} \,d\chi' \int_{0}^{\chi_{\beta}} \,d\chi'' \int \,\frac{d^2\vec{l'''}}{(2\pi)^2}W(\chi,\chi_{\alpha})W(\chi',\chi_{\beta})W(\chi'',\chi_{\beta})\frac{d\overline{\tau}}{d\chi''}
     \cos(\omega) \frac{9\Omega_{m}^2H_{0}^4}{4a^2k^2}\langle \widetilde{\delta}_{m}(\vec{l};\chi)\widetilde{\delta}_{m}(\vec{l'''};\chi')\widetilde{\delta}_{d}(\vec{l''} - \vec{l'''};\chi'') \rangle ,\\
\label{eq: reducedshear5}
     =(2\pi)^2 \delta_{D}(\vec{l} + \vec{l''}) \int_{0}^{\chi_{\alpha}} \,d\chi \int \,\frac{d^2\vec{l'''}}{(2\pi)^2}W(\chi,\chi_{\alpha})W(\chi,\chi_{\beta})\frac{d\overline{\tau}}{d\chi}
     \cos(\omega) \frac{9\Omega_{m}^2H_{0}^4}{4a^2}B_{\delta_{m}\delta_{m}\delta_{d}}(\vec{l}/\chi,\vec{l'''}/\chi,(\vec{l''} - \vec{l'''})/\chi;\chi),
\end{gather}
where $\omega \equiv 2\phi_{l''} - 2\phi_{l'''}$ , $\delta_D$ is the Dirac delta function, and $B_{\delta_{m}\delta_{m}\delta_{d}}$ is the matter-matter-dust bispectrum. We also used Poisson's equation to relate the density fluctuations ($\delta_{m}$) to the gravitational potential ($\phi$).

\begin{center}
\begin{figure}[h!]
\includegraphics[scale=0.7]{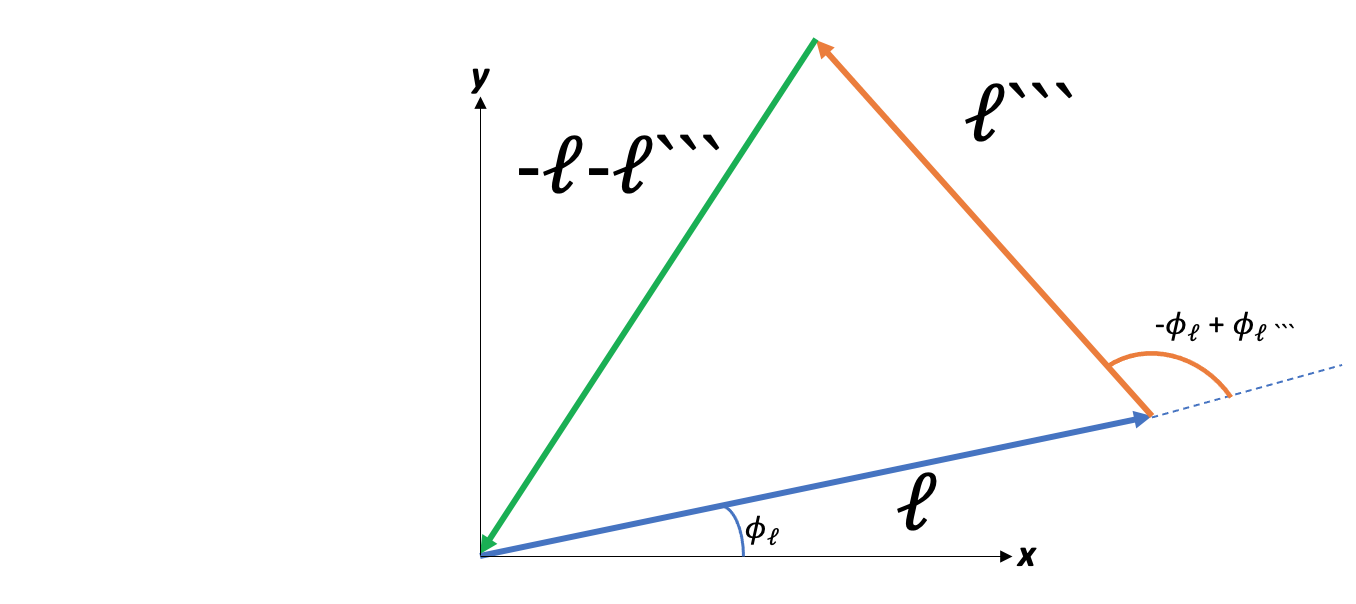}
\caption{Triangular configuration of $\vec{l}$ modes that give non-zero contributions to the bispectrum.\label{fig:kTriangle}}
\end{figure}
\end{center}

The bispectrum physically describes the contribution to the skewness of the local density from each triplet of Fourier modes. It is the next-order statistic after the power spectrum; in general, multiplication of fields such as the operation in Eq.~(\ref{eq: wt_reduced_shear_real}) mixes higher-order statistics into lower-order ones. Figure~\ref{fig:kTriangle} gives a geometric image of how the $\vec{l}$ modes must be aligned to contribute to the power spectrum correction in Eq.~(\ref{eq: reducedshear5}).

Now that we have defined computed $\Delta C_{ll'}^{\alpha \beta}$, we must compute the matter-matter-dust bispectrum. Since the bispectrum is a next level statistic, we will need a model that describes the distribution of matter and dust. We will use the \textit{Halo Model} approach to compute the bispectrum.

\subsection{Halo Model Approach}

The halo model approach provides an analytic formulation of the clustering of dark matter and galaxies under the assumption that all dark matter is in the form of virialized halos \citep[e.g.][]{2000MNRAS.318..203S}. Halos form with a distribution of masses as described in the \citet{1974ApJ...187..425P} formalism, and are biased with respect to the matter density fluctuations \citep{1984ApJ...284L...9K, 1989MNRAS.237.1127C}. Modern halo model approaches use mass functions and biases fit to $N$-body simulations; here we use the \citet{1999MNRAS.308..119S} model. Given a mass profile for the halo, a mass function, and spatial distribution, we can compute the necessary power spectra. This approach has proven useful not just for galaxy counts, but can also be used for diffuse signals within haloes such as the Sunyaev-Zel'dovich effect \citep[e.g.][]{1999ApJ...526L...1K, 2002MNRAS.336.1256K}; here we apply it to circumgalactic dust.

We start by first defining a set of ingredients that are needed to construct the halos. We need to specify a matter profile for the halo, an abundance (the halo mass function, $dn/dM$), and the spatial distribution of halos. If we follow the set up as in \cite{2001ApJ...548....7C}, then we can define the following ingredients as 
\begin{equation}
    \label{eq: halo_profile}
    \rho(r,M,z) = \frac{\rho_{c}(z,M,M_{crit})}{[r/r_{c}(z,M,M_{crit})][1 + r/r_{c}(z,M,M_{crit})]^{2}},
\end{equation}
where $M_{crit}$, $r_{c}$, and $\rho_{c}$ are the critical mass scale of the halos, and the characteristic radius and density, respectively, defined in \cite{2001ApJ...548....7C} in their Eq. (20) - Eq. (22).
We then define the normalized Fourier transform of the density
\begin{equation}
    \label{eq: fourier_transform_halo_profile}
    y(k,M,z) = \frac1M \int^{r_v}_{0} 4\pi r^2 \rho(r,M,z) \frac{\sin(kr)}{kr} dr,
\end{equation}
where $M$ is the total mass of the halo. We follow the same definition of the halo mass function and all its components in \cite{2000MNRAS.318..203S}. We will also follow \cite{2001ApJ...548....7C} in describing halo clustering in terms of the halo bias coefficients: 
\begin{eqnarray}
b_0(z,M) &\equiv& 1, \nonumber \\
b_{1}(z,M) &=& 1 + \frac{a\nu(z,M) - 1}{\delta_c} + \frac{2p}{\delta_c(\left[a\nu(z,M)\right]^p + 1)},
~~~{\rm and}
\nonumber \\
b_2(z,M) &=& \frac{8}{21}\left[ b_1(z,M) - 1 \right] + \frac{\nu(z,M) - 3}{\sigma(M)^2} + \frac{2p\left[ 2p + 2a\nu(z,M) - 1 \right]}{\delta_c^2(\left[a\nu(z,M)\right]^p + 1)},
\label{eq: bias_1}
\end{eqnarray}
where $\nu \equiv \left[ \delta_c/\sigma(M) \right]^2$, $\delta_c = 1.68$ is the of overdensity for spherical collapse in the Einstein de-Sitter model (which we assume in this work), $\sigma(M)$ is the root-mean-square (rms) of fluctuations in spheres containing mass M at the virial radius, and $(a,p) = (0.707,0.3)$, respectively.

Once these ingredients are defined, we are able to construct the halo model bispectrum. The basic idea in constructing this bispectrum is that we want to compute the three point correlation function (the inverse Fourier transform of the bispectrum) for three points located in any permutation of three halos. This means there are three possible contributions to the bispectrum: single, double, and triple halo contributions. The triple halo contribution physically represent the case when a single particle occupies each halo. The double halo represents the case when two particles are in the same halo and the third in a different halo. And the single halo represents the case when three particles are in one halo.

We now want to calculate the contributions from the single, double, and triple halo configurations. It will be useful to define the following integrals:
\begin{equation}
    \label{equ: I-integrals}
    I^\mu_{\beta}(k,z) = \int_{M_{min}}^{M_{max}} dM \left( \frac{M}{\rho_{b}}\right)^{\mu}\frac{dn}{dM} b_{\beta}(M,z) \prod_{i=1}^{\mu} y_{i}(k_{i},M,z),
\end{equation}
where $\rho_{b} = \Omega_{m}\rho_{critical}$ is the background matter density; $\mu$ indicates the number of points in the correlation function that are in that halo (a minimum of $\mu=1$, and a maximum of $\mu=N$ for the 1-halo term in an $N^{\rm th}$ order correlation function); $\beta$ is the order of the bias coefficient for that halo; and $M_{min}$ and $M_{max}$ are the minimum and maximum halo masses of interest. The different contributions will be defined as different combinations of these integrals. 

The total halo bispectrum, $B_{t}$, can be written following the notation of \citet{2001ApJ...548....7C} as 
\begin{equation}
    \label{equ: B_t}
    B_{t} = B_{PPP} + B_{Phh} + B_{hhh},
\end{equation}
where the single halo contribution is
\begin{equation}
    \label{equ: B_PPP}
    B_{PPP}(k_{1},k_{2},k_{3}) = I^{3}_{0}(k_{1},k_{2},k_{3}),
\end{equation}
the double halo contribution is
\begin{equation}
    \label{equ: B_Phh}
    B_{Phh}(k_{1},k_{2},k_{3}) = I^{2}_{1}(k_{1},k_{2})I^{1}_{1}(k_{3})P^{Lin}(k_{3}) + Perm.,
\end{equation}
and the triple halo contribution is
\begin{equation}
    \label{equ: B_hhh}
    B_{hhh}(k_{1},k_{2},k_{3}) = \left(2F_{2}(\vec{k}_{1},\vec{k}_{2})I^{1}_{1}(k_{3}) + I^{1}_{2}(k_{3}) \right)I^{1}_{1}(k_{1})I^{1}_{1}(k_{2})P^{Lin}(k_{1})P^{Lin}(k_{2}) + Perm.,
\end{equation}
where $F_{2}(\vec{k}_{i},\vec{k}_{j})$ is the two point kernel from perturbation theory and the $Perm.$ is the permutation of $k_{1}$, $k_{2}$, and $k_{3}$. The ``P'' and ``h'' labels indicate Poisson-like and halo clustering contributions.

\subsection{Building the Dust Model Bispectrum}
\label{subsec: dust_bispectrum}

We can carry out the exact same procedure to calculate the matter-matter-dust bispectrum as in the halo model. The only difference is now rather than having three connected points occupied by a matter contribution (distributions defined Eq. (\ref{eq: halo_profile}) and Eq. (\ref{eq: fourier_transform_halo_profile})), there will be a contribution from dust. This means we will need to define a dust density profile and the total amount of dust within each halo. 
We calculate the density profile of the dust by using the extinction model that was measured in \cite{2010MNRAS.405.1025M} in Eq. (30). First, we need to understand the geometry of the measurement. Since extinction is measured along some line of sight through a dusty halo, it is measuring the amount of extinction due to a column of dust. But we can relate this to the true geometry of the dust, which is typically an annulus with a galaxy at the center. This equivalent to solving the integral equation
\begin{equation}
\label{eq:2d-3d_integral}
\begin{split}
\Sigma(R) = \int_{-\infty}^{\infty} \rho(r)dr_{\parallel},\\
r = \sqrt{R^2 + r_{\parallel}^2}
\end{split}
\end{equation}
given the surface density profile, $\Sigma(R)$, and solving for the volumetric density, $\rho(r)$ and $r_{\parallel}$ is the radial distance along the line of sight. For the case of a spherical geometry and that $\rho(r)$ is a power law, which is the case for our purposes, $\Sigma(R)$ is also a power law. Since Eq. (30) in \cite{2010MNRAS.405.1025M} is a power law, with index $\gamma = 0.84$, then the dust density must also be a power law. This gives a density profile
\begin{equation}
\label{eq:dust_density}
\begin{split}
\rho_{d}(r) = \rho_{norm} r^{-1-\gamma},\\
\rho_{norm} = \frac{M_{d}}{4 \pi \int_{r_{eff}}^{r_{vir}}r^{-1-\gamma}r^2dr},\\
\overline{\rho}_{d}(z) = \int M_{d}(z,M)\frac{dn}{dM}dM,
\end{split}
\end{equation}
where $r_{eff}$ is the effective radius that encompasses the central galaxy and $r_{vir}$ is the virial radius of the halo and $\overline{\rho}_{d}(z)$ is the mean density of dust in the Universe for a given dust mass, $M_{d}$. Using Eq. (\ref{eq:dust_density}), we can define the mean optical depth per unit length
\begin{equation}
\label{eq: mean_tau}
 \begin{split}
    \frac{d\overline{\tau}}{d\chi} = \kappa_{d}(1+z)^2\overline{\rho}_{d}(\chi),\\
   \frac{d\overline{\tau}}{dz} = \kappa_{d}(1+z)^2\overline{\rho}_{d}(z) \frac{d\chi}{dz},
\end{split}   
\end{equation}
where we rewrite our mean optical depth as function of redshift instead of $\chi$.

Once we define the dust mass and density profile of the dust, we can follow the same mathematical set up as in Eq. (\ref{eq: fourier_transform_halo_profile}) - (\ref{equ: B_hhh}) to build the matter-matter-dust bispectrum. Define one of the $k_{i}$ as the contribution due to dust instead of matter, this means using the density profile of the dust whenever the argument corresponds the designated $k_{i}$. The new equations become
\begin{align}
    u(k,M,z) &= \frac1{M_{d}(M)}\int_{0}^{r_{v}}4\pi r^2\rho_{d}(r,M_{d}(M),z)\frac{\sin{(kr)}}{kr}dr \nonumber \\
    &= \frac{4\pi r_{v}^{1.16}\rho_{norm}}{M_{d}(M,z)}  0.86 \times {_1}F_{2}\left(0.58,1.5,1.58,-\frac14 (k r_{v})^2\right),
\label{eq: dust_profile_fourier}
\end{align}

\begin{align}
\label{eq: I03_dust}
    I^{3,dust}_{0}(k_{1},k_{2},k_{3},z) &= \int^{M_{max}}_{M_{min}}dM\left(\frac{M}{\rho_{b}}\right)^2\left(\frac{M_{d}}{\overline{\rho}_{d}}\right)\frac{dn}{dM}y(k_{1},M,z)y(k_{2},M,z)u(k_{3},M,z),\\
\label{eq: I12_dust}
    I^{2,dust}_{1}(k_{i},k_{j},z) &= \int^{M_{max}}_{M_{min}}dM\frac{dn}{dM}b_{1}(z,M)f(k_{i},M,z)f(k_{j},M,z),\\
    I^{1,dust}_{1}(k_{i},z) &= \int^{M_{max}}_{M_{min}}dM\frac{dn}{dM}b_{1}(z,M)f(k_{i},M,z),~~{\rm and}\\
    I^{1,dust}_{2}(k_{i},z) &= \int^{M_{max}}_{M_{min}}dM\frac{dn}{dM}b_{2}(z,M)f(k_{i},M,z),
\end{align}
where $f(k_{i},M,z) = \frac{M}{\rho_{b}}y(k_{i},M,z)$ for $i=1,2$ and $f(k_{i},M,z) = \frac{M_{d}}{\overline{\rho}_{d}}u(k_{i},M,z)$ for $i=3$, and $_{1}F_{2}$ is the generalized hypergeometric function. Now, the final piece of information we require is the total mass of dust within the circumgalatic halo.

\section{Building a model of Dust in Galatic Halos}
\label{sec:dust_measurement}

We now need to build a model for the amount of halo dust as a function of halo mass and redshift. Dust may be present at all radii in a halo, from near the center out to the virial radius, but for the purposes of this study, we are interested in dust in the circumgalatic halo. The minimum (projected) radius that we are interested in is set by the minimum radius in a halo where we might select a background galaxy as a source. At very small separations, the background source and the source galaxy overlap (``{blending}''). Both biases due to exclusion of galaxies by blending and due to neighbor effects on second moment measurement are important in modern weak lensing analyses \citep[e.g.][]{2015MNRAS.449.1259S, 2016ApJ...816...11D, 2018A&C....24..129M, 2021MNRAS.503.4964G}, but they are outside the scope of this paper. In this section, we will investigate different models that relate the dust mass to the stellar mass.

Measurements of circumgalactic dust are limited, and so we aim here to construct a simple model, bounded by physical constraints, with a few free parameters tuned to match the observations of \citet{2010MNRAS.405.1025M}. This does not allow a determination of the redshift dependence, but fortunately the central redshift of $z=0.36$ in \citet{2010MNRAS.405.1025M} is close to the midpoint of the line-of-sight integral for background galaxies at $z\sim 1$. In order to be consistent with \citet{2010MNRAS.405.1025M}, we used the $R_{V} = 3.1$ dust extinction model from \citet{2001ApJ...548..296W}.

We build an upper bound case (most possible dust) for the scenario of dust ejected from the central galaxy by scaling from the stellar mass of the central galaxy. For the stellar mass-to-halo mass relation model $M_{\rm stellar}(M)$, we use Eq.~(3) in \cite{2013ApJ...770...57B} with intrinsic parameters defined in Section 5. The upper bound case for the dust mass model would be to assume the dust mass scales as the stellar mass within a galaxy via $M_{d,upper} = yM_{stellar}(M,z)$, where $y \sim 0.015$ \citep{2016MNRAS.455.4183V} is the yield of dust-forming elements produced through stellar evolution and $M$ is the mass of the halo. (The precise choice of $y$ is not important since the normalization parameter for our base model (Sec. \ref{subsec: base_model}), $\alpha$, is degenerate with $y$ and absorbs any error.) Our assumption for the yield in this dust model is the total yield of metals goes to dust, which is an overestimate as we know not all metals are ejected into the circumgalactic region, and not all of the ejected metals necessarily go to dust formation.

It is not {\slshape a priori} clear how far from this upper bound we should expect the true dust mass to be. \citet{2011MNRAS.412.1059Z} show through hydrodynamic simulations that, depending on transport mechanism, that dust can found out about 10 $h^{-1}$Mpc from the galactic center. In their model they use the data from \cite{2010MNRAS.405.1025M} in order to constrain their hydrodynamic simulations, which was the result of the smooth particle hydrodynamics (SPH) code \textit{GADGET}. In a similar study, though not using data from \cite{2010MNRAS.405.1025M}, we see in \cite{2016MNRAS.457.3775M} that the dust-metal ratio is typically higher out in the circumgalatic medium of a halo. Their simulation was executed using the moving mesh code \textit{AREPO}.

\subsection{Base Model}
\label{subsec: base_model}
Since there are few measurements of dust out into virial radius of the halos, and the \cite{2010MNRAS.405.1025M} data are sufficient to constrain two parameters (an amplitude and power law slope), we make a two-parameter model for $M_d(M)$:
\begin{align}
\label{eq: dust_model}
    M_{d}(M,z) = \alpha M_{d,upper}(M,z)e^{-M/M_{cutoff}},
\end{align}
where $\alpha$ is a dimensionless overall scaling and $M_{cutoff}$ is a cutoff mass. A cutoff may be physically motivated, e.g., by grain evaporation in high-temperature haloes, but in this paper we leave the cutoff scale to be determined empirically. As noted above, the value of the yield parameter, $y$, defined in $M_{dust,upper}$ does not affect our base model since the parameter $\alpha$ normalizes our amplitude to the measured extinction.

We can determine the parameters in Eq. (\ref{eq: dust_model}) by relating them to the \cite{2010MNRAS.405.1025M} measurement at two scales. At large scales, the two-halo term dominates, and we can constrain a combination of the dust density and dust bias. We relate Eq.~(30) in \cite{2010MNRAS.405.1025M} to the two halo quantity via
\begin{align}
\label{eq: twohalo_amp_measure1}
    \langle A_{V} \rangle (r_{p}) &= b_{g}b_{d}\frac{d\tau}{dr_{\parallel}}\frac{2.5}{\ln{10}}\int_{-\infty}^{\infty} \xi(r_{p},r_{\parallel})dr_{\parallel} \\
    \label{eq: twohalo_amp_measure2}
    &= b_{g}b_{d}\frac{d\tau}{dr_{\parallel}}\frac{2.5}{\ln{10}}\frac{1}{2 \pi}\int_{0}^{\infty}k_{p}dk_{p}P(k_{p})J_{0}(k_{p}r_{p}),
\end{align}
where $b_{g}$ and $b_{d}$ are the galaxy and dust bias functions, $d\tau/dr_{\parallel}$ is the change in optical depth along the line of sight, $P(k_{p})$ is the nonlinear matter power spectrum, $J_{0}(k_{p}r_{p})$ is the Bessel function of the first kind, and $k_{p}$ and $r_{p}$ are the wave vector and radius perpendicular from the line of sight. We use this equation at $r_p\sim 3h^{-1}$Mpc (well into the two-halo regime, and where there is still a detection in \citealt{2010MNRAS.405.1025M}).

We can break down the meaning of Eq. (\ref{eq: twohalo_amp_measure1}) by examining each piece of the expression. The left hand side is an average over realizations of extinction measurements of different galaxies due to dust in the V-band. The right hand side is composed of a projected 3D matter power spectrum onto the line of sight (integral) but is then converted to a galaxy-dust power spectrum with the bias coefficients ($b_{g,d}d\tau/dr_{\parallel}$) (\citealt{2003MNRAS.340..580T}). The numerical coefficient is necessary to convert measurements in magnitudes to optical depth measurements (extinction). 

We can further constrain our model by using the one-halo term, which constrains how much dust would be in a halo the size of the one used in \cite{2010MNRAS.405.1025M}, i.e., the halo hosting a galaxy of luminosity $0.5L_*$. We can compute the total mass of dust in this sphere by integrating the volumetric extinction profile (Eq.~\ref{eq:2d-3d_integral}) over the volume of a sphere at the virial radius of the halo.

For our analysis, we take the galaxy bias to be the halo bias corresponding to the galaxy sample in \cite{2010MNRAS.405.1025M} (see below).
Now the term in Eq. (\ref{eq: twohalo_amp_measure2}), $b_{d}d\tau/dr_{\parallel}$ is related to the dust mass function by
\begin{align}
\label{eq: twohalo_amp_theory}
    b_{d}\frac{d\tau}{dr_{\parallel}} &= \kappa_{d}(1+z)^2\int_{0}^{M_{max}}dM b_{1}(M)\frac{dn}{dM}M_{d}(M).
\end{align}
We now have two unknowns ($\alpha$ and $M_{cutoff}$) and two constraints (one and two-halo term), therefore, we can uniquely determine the parameters in Eq.~(\ref{eq: dust_model}). To do this, we rewrite $\alpha$ as a function of $M_{cutoff}$ in Eq. (\ref{eq: dust_model}) then compute $b_{d}d\tau/dr_{\parallel}$ in Eq. (\ref{eq: twohalo_amp_measure2}) using the radius of the halo relevant to the dust mass measurement in \cite{2010MNRAS.405.1025M}. However, they claim that for an isothermal sphere model for the halo of a galaxy of luminosity $0.5L_*$, that $r_{\rm virial} \sim 110h^{-1}\,$kpc. But based on \cite{2006MNRAS.368..715M}, we see that a halo with that luminosity has a central halo mass of M $\sim$ $4.11\times 10^{11} h^{-1}M_{\odot}$, which leads to a value of $r_{p} \sim$ 177 $h^{-1}\,$kpc. With this new radius, we compute the mass of dust measured to be about $4.1\times 10^{7}h^{-1}M_{\odot}$. Now we can compute $M_{cutoff}$ by finding the value that makes the two equations for $b_{d}d\tau/dr_{\parallel}$ (the equation that relates $b_{d}d\tau/dr_{\parallel}$ to the observed extinction, Eq.~\ref{eq: twohalo_amp_measure2}, and the halo model in Eq.~\ref{eq: twohalo_amp_theory}) agree. This value turns out to be $6.73\times 10^{12} h^{-1}M_{\odot}$ which gives a value of 0.353 for $\alpha$ and completely constrains our model.

\subsection{Ceiling and Floor Models}

We will also define two ``extreme'' cases for our dust model that are within physical bounds of dust production/destruction. These extreme models will give us an idea of how sensitive our analysis is to the extrapolation of our dust mass model with redshift. We will define upper and lower bounds to our fiducial model (\textit{base} model) based on the current understanding of circumgalactic dust and its evolution in order to test the sensitivity of our analysis to the types of dust mass models.

The essence of our argument for our upper bound (\textit{ceiling} model) is based on the stellar metallicity relation at $z \sim 2$, which is Eq. (4) in \cite{2021arXiv210906044K}
\begin{equation}
\label{eq:stellar_metal}
\log\left( Z_{stellar}/Z_{\odot} \right) = -0.81 + 0.32 \log \left( M_{stellar} / 10^{10}M_{\odot} \right),
\end{equation}
where $Z_{stellar,\odot}$ are the stellar and solar metallicities, respectively. If we assume that total available yield of metals go like $M_{d,upper} = 0.015M_{stellar}$ and that $Z_{\odot} = 0.0142$, then we can find the fraction of metals in stars by dividing the mass of metals in stars (Eq. (\ref{eq:stellar_metal}) * $M_{stellar}$) with the total available metals. We also take into account that certain metals (Ar, Ne) don't form dust at all \cite{2019arXiv191200844L}, which will reduce the amount of dust by about $\sim 15\%$. Since we are assuming the dust either goes into stars or circumgalatic dust, then the amount of metals that would go into circumgalatic dust would be given by
\begin{equation}
\label{eq:circum_dust}
M_{d}(z = 2.0,M)/M_{d,upper} = 0.85 \times \left(1 - 0.147\times \log \left( \frac{M_{stellar}(z=2.0,M)}{10^{10}M_{\odot}} \right)^{0.32} \right).
\end{equation}
We then linearly interpolated this dust model with our base model at redshift 0.36 to give
\begin{equation}
\label{eq:dust_model_interp}
M_{d}(z,M) = M_{d}(z=0.36,M) - \left(\frac{z - 0.36}{2.0 - 0.36}\right) \times (M_{d}(z=0.36,M) - M_{d}(z=2.0,M)).
\end{equation}
For our lower bound (\textit{floor} model), we make the assumption that no dust contributes to an shear measurements past redshift of 0.8, which is the extent of the \cite{2010MNRAS.405.1025M} study.

\section{Results}
\label{sec:results}

We evaluated the power spectra and corrections for all 55 redshift pairs from the set of 10 bins centered at $\{z_{i}\}=\{0.1,0.3,0.5,...,1.9\}$, which is similar to the redshift bins used in the \textit{Roman} survey forecast. All of the cross- and auto-power spectra are incorporated in the Fisher matrix, but for display purposes we focus on the auto-correlations. We used the values $(M_{cutoff},\alpha) = (6.72\times 10^{12} h^{-1}M_{\odot},0.353)$ computed in the previous section based on extinction measurements of dust. All of our matter power spectra, $P(k)$, were computed from {\sc CLASS} \citep{2011JCAP...07..034B} and we used the default {\sc HMcode} prescription for the non-linear power spectrum \citep{2015MNRAS.454.1958M}.

%In the left panel of Fig. (\ref{fig:multiplot_integrand_dustmass}), we plot the integrand of Eq. (\ref{eq: twohalo_amp_theory}) multiplied by $M$. This allows us to see which mass ranges contributed the most to the two halo amplitude. We can see that halo masses $\sim 10^{12}h^{-1}M_{\odot}$ contributed the most to the two halo term.

\begin{figure}[h!]
\centering{\includegraphics[width=6.75in]{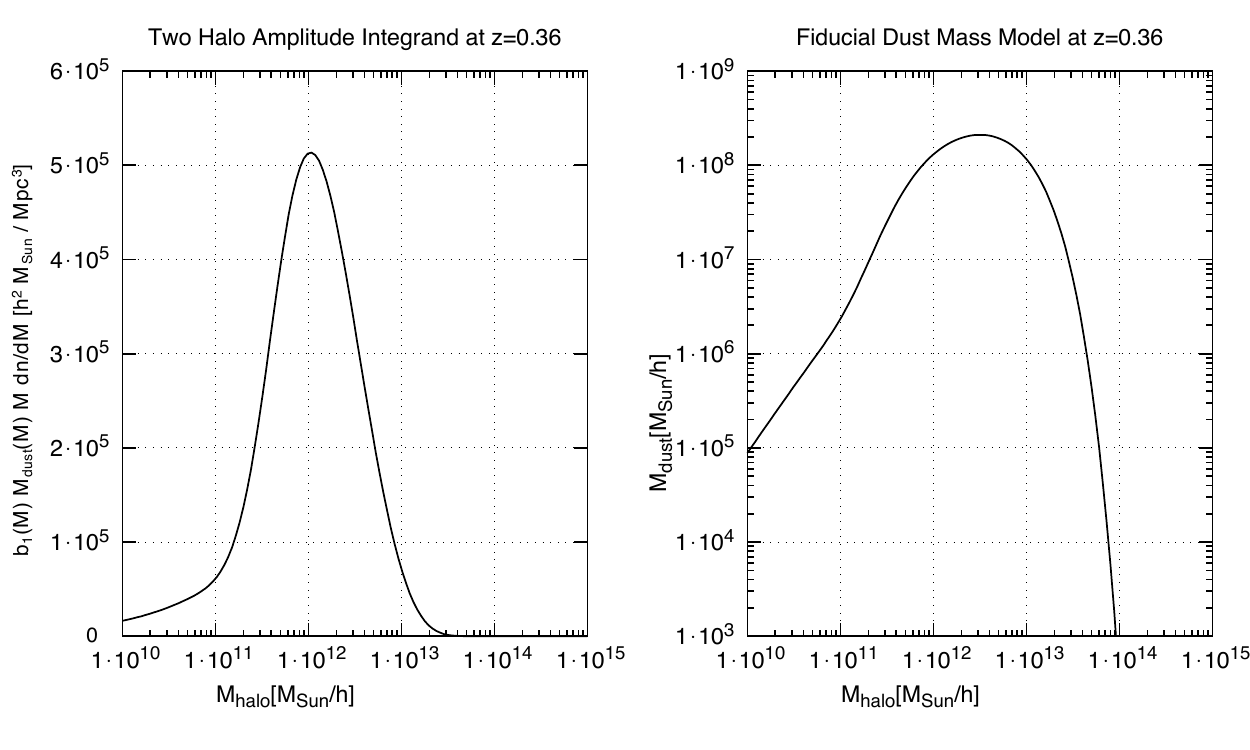}}
\caption{The integrand of Eq. (\ref{eq: twohalo_amp_theory}), multiplied by $M$ so that on semilog scale the contribution to the integral is proportional to the area in the plot (left panel), shows that most of the contributing mass range is around $\sim 10^{12}h^{-1}M_{\odot}$. The plot of the fiducial dust mass model (right) shows a rapid decrease in dust mass for halos greater than $\sim 3\times 10^{13}h^{-1}M_{\odot}$.\label{fig:multiplot_integrand_dustmass}}
\end{figure}

\begin{figure}[h!]
\plotone{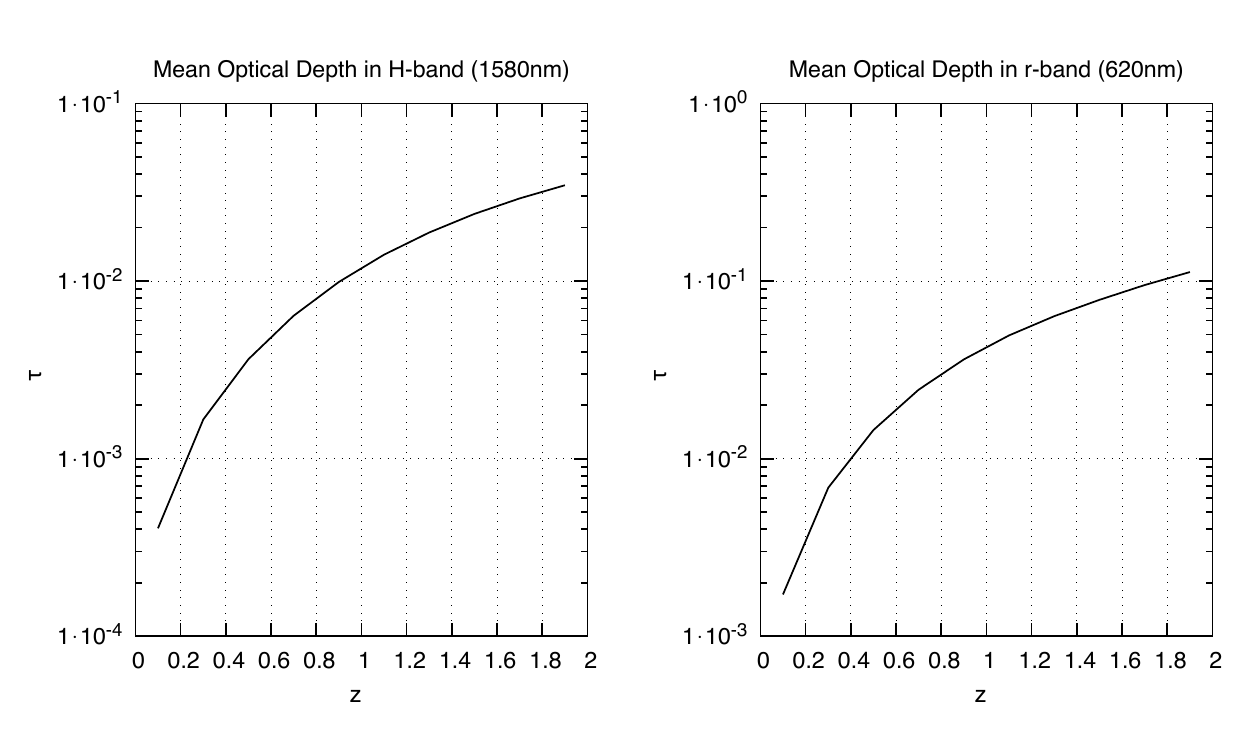}
\caption{The mean optical depth of dust in the H-band (left) and r-band (right) in the observer frame, integrated along the line of sight from 0 to $z$. The increase in optical depth for increasing redshift is due dust strongly interacting with bluer wavelengths of light.\label{fig:multiplot_taudust}}
\end{figure}

In Fig.~\ref{fig:multiplot_integrand_dustmass}, we show the fiducial dust mass at $z=0.36$. The left panel shows the integrand of Eq. (\ref{eq: twohalo_amp_theory}) multiplied by $M$, i.e., the contribution to the bias-weighted dust optical depth per logarithmic range in $M$. This allows us to see which mass ranges contributed the most to the two halo amplitude. We see that halo masses $\sim 10^{12}h^{-1}M_{\odot}$ contribute the most to the two halo term. In the right panel of Fig.~\ref{fig:multiplot_integrand_dustmass}, we see that the dust mass falls off quickly for $M > 10^{13}h^{-1}M_{\odot}$. In Fig.~\ref{fig:multiplot_taudust}, we show the mean optical depth obtained by integrating Eq. (\ref{eq: mean_tau}) up to some redshift $z$ (this is comparable to Fig.~9 of \citealt{2010MNRAS.405.1025M}). As expected, the mean optical depth increases as the redshift increases both because of the path length but also because of the bluer rest-frame wavelength.

\begin{figure}
    \centering
    \includegraphics[scale = 1.2]{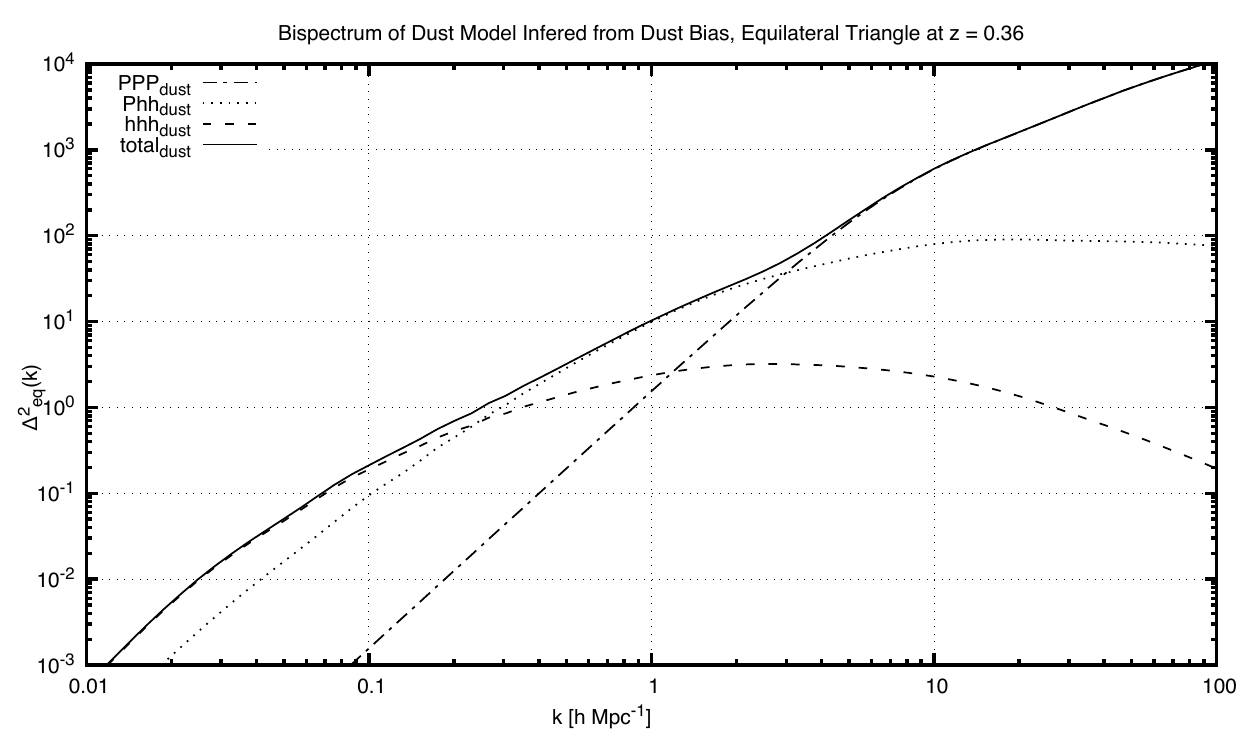}
    \caption{The single, double, and triple halo contributions to our dust bispectrum for a equilateral k-triangle configuration at redshift $z=0.36$. We compute the scaled dust bispectrum defined by $\Delta^2_{eq}(k) \equiv k^3/2\pi^2\sqrt{B_{t}(k)}$ for an equilateral $k$-triangle configuration. As is usual for bispectra, the three- ($hhh_{dust}$) and two-halo ($Phh_{dust}$) terms are the dominant contributions to the total bispectrum (total$_{\rm dust}$) at large scales, while the one-halo term $PPP_{dust}$ becomes dominant at smaller scales.}
    \label{fig: dust_bispec}
\end{figure}

The bispectrum and its components for the base model are plotted in Fig.~\ref{fig: dust_bispec}. We can see that for large scales (small $k$), the triple and double halo terms ($hhh_{dust}$ and $Phh_{dust}$) dominate the total bispectrum, while at smaller scales the single halo term ($PPP_{dust}$) is most important. Physically, this makes sense since we expect correlations involving different halos to be important at large scales and correlations within the same halo to be more important at small scales. The transition to the single halo term dominating occurs at $k\sim 3h\,$Mpc$^{-1}$, which is a smaller scale than for the matter bispectrum \citep[e.g.][Fig.~1]{2001ApJ...548....7C}, because the dust-to-total-mass ratio falls off for higher halo masses; thus while the matter bispectrum contains a large contribution from the 1-halo term for massive ($\gtrsim 10^{14}\, M_\odot$) halos, the matter-matter-dust bispectrum does not.

\begin{figure}[h!]
\centering{\includegraphics[width=5.5in]{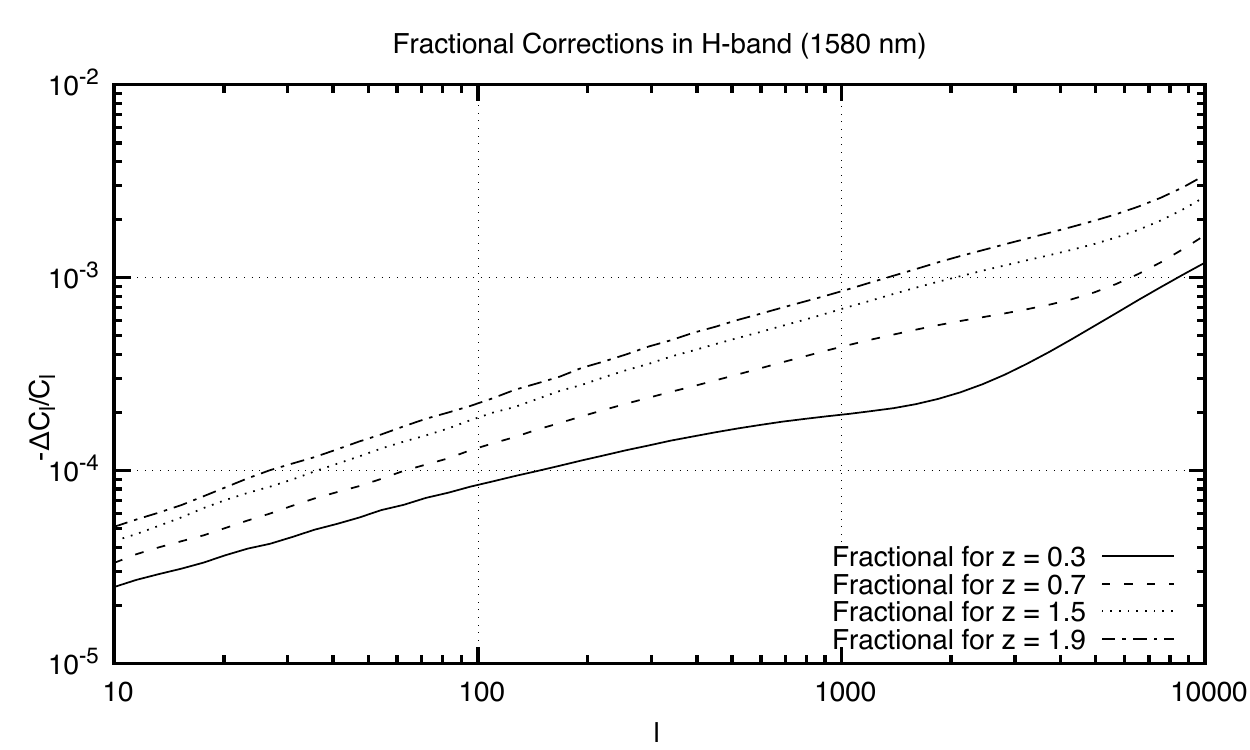}
\includegraphics[width=5.5in]{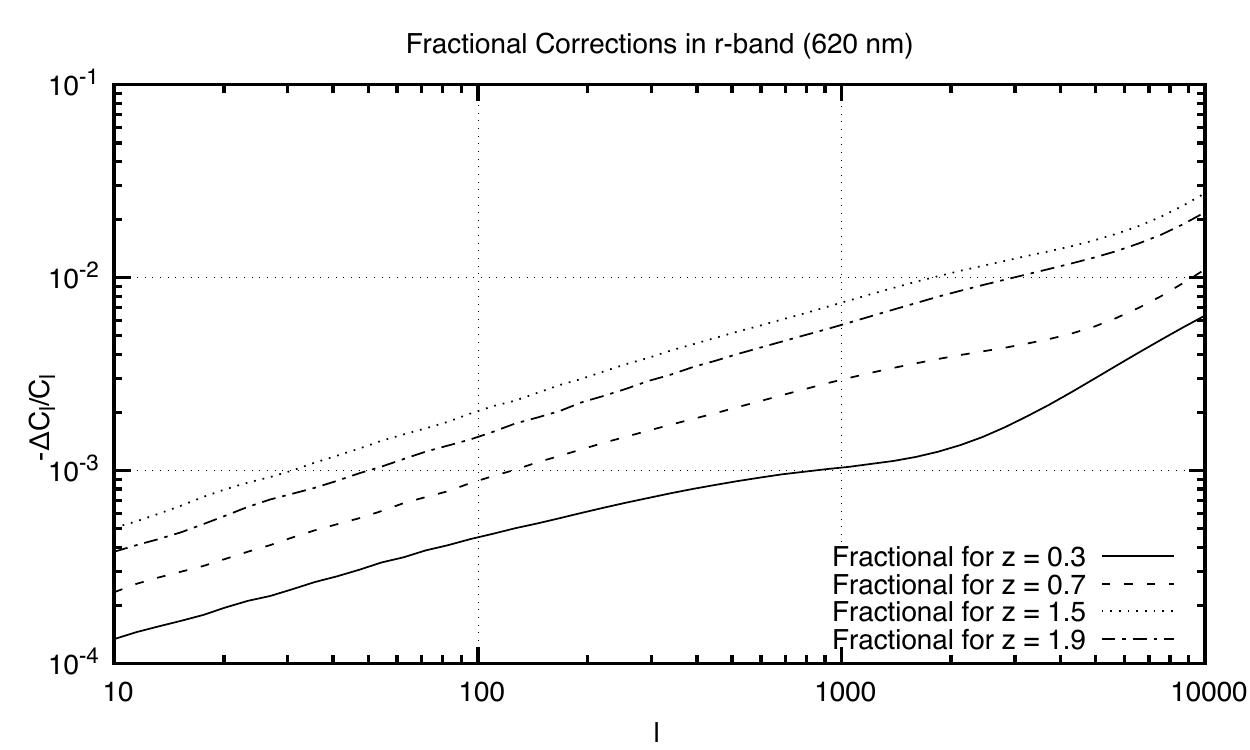}
}
\caption{The fractional corrections to the shear auto-power spectra, $-\Delta C_l/C_l$, for the indicated redshift slices. The top panel shows the observer-frame H-band, and the bottom panel shows the observer-frame r-band.
%ratio of the corrections with the shear power  spectrum  in the H-band (observer frame) evaluated at the same redshift slice.
\label{fig:fractional_h}}
\end{figure}

\begin{figure}[h!]
\centering{
\includegraphics[width=5.5in]{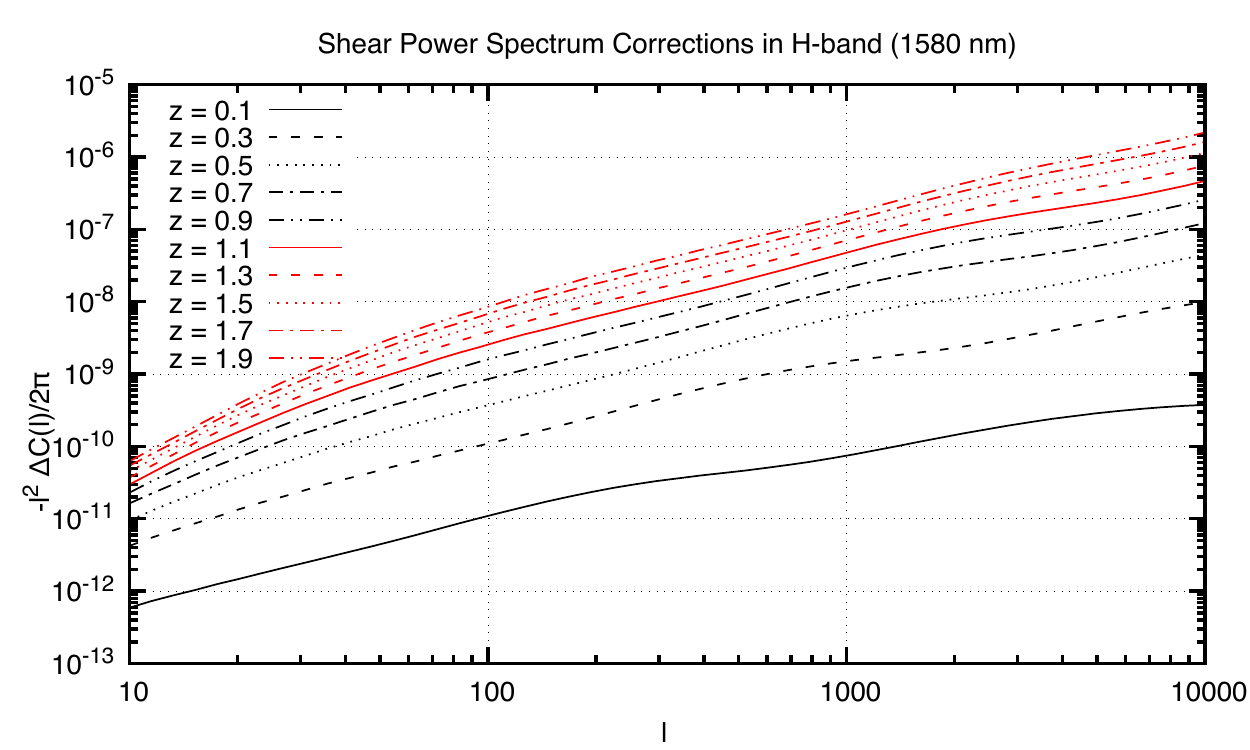}
\includegraphics[width=5.5in]{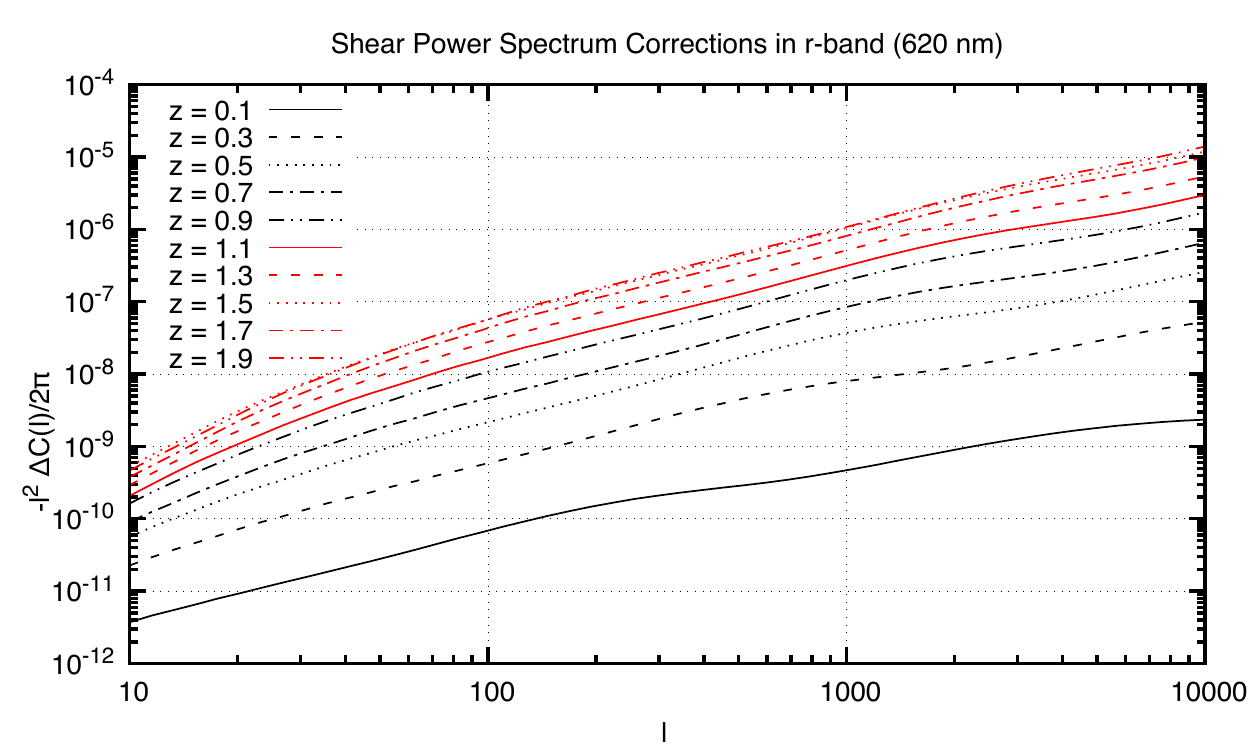}
}
\caption{The corrections $-l^2\Delta C_l/2\pi$ to the E-mode shear auto-power spectra. The top panel shows the observer-frame H-band (\textit{Roman}, 1580 nm) and the bottom panel shows the observer-frame r-band (\textit{LSST}, 620 nm).
%Third order corrections to the E-mode shear power spectrum in the H-band (observer frame) of \textit{Roman} (1580 nm) evaluated at the same redshift slice.
\label{fig:reducedshear_h}}
\end{figure}

The correction $\Delta C_l$ depends on the slope of the number counts, $d\ln N/d\ln F$; we used the results for the {\slshape Roman} reference survey \citep{2021MNRAS.507.1746E}. The derivative was computed by finite difference, varying the effective flux limit by varying the number of exposures in the {\slshape Roman} Exposure Time Calculator \citep{2013ascl.soft11012H}.
We computed the ratio $\Delta C_l/C_l$ of the correction to the shear power spectrum relative to its base value. We found that their ratios deviated very slightly across a range of redshift samples as displayed by the black curves in Fig.~\ref{fig:fractional_h}. We see in Fig.~\ref{fig:fractional_h} that the maximum fractional correction to the power spectrum $|\Delta C_l|/C_l$ is about $3\times 10^{-3}$ (H band) or $2\times 10^{-2}$ (r band). We can see in the bottom panel of Fig.~\ref{fig:fractional_h} (r band) that the fractional correction $|\Delta C_l|/C_l$ is slightly greater at $z=1.5$ than at $z=1.9$. The absolute correction $|\Delta C_l|$ is still increasing with redshift as we would expect (bottom panel of Fig. \ref{fig:reducedshear_h}), just not as fast as $C_l$. This follows the same reasoning as the increase in mean optical depth for increasing redshift since the integrand in Eq.~(\ref{eq: reducedshear5}) is positively increasing with mean optical depth. The corrections to the auto-power spectra are shown in Fig.~\ref{fig:fractional_h} (showing fractional corrections) and Fig.~\ref{fig:reducedshear_h} (showing absolute corrections $-l^2\Delta C_l/2\pi$).

The change in power spectrum is a complicated function of scale and redshift. In order to interpret our results, and understand their impact on weak lensing surveys, we apply several error metrics. The first ($Z^2$) operates purely in data vector space and compares the change in power spectrum to statistical errors. The second investigates how this change in the data vector would affect a measurement of the amplitude of structure ($\sigma_8$ with other parameters held fixed). Our error metrics depend on the survey parameters; here we use the {\slshape Roman} 2000 deg$^2$ reference survey, as described in \citet{2021MNRAS.501.2044T}, restricted to the 10 bins at $z<2$ (there are higher redshift bins, but they have few galaxies and as we push to higher redshifts the extrapolation of the dust model used in this paper becomes extreme).

The first metric we employ is the $Z^{2}$ error metric (used for {\slshape Roman} error budgeting, \citealt{2021MNRAS.501.2044T}; see \citealt{2013MNRAS.429..661M, 2018arXiv180901669T, 2020A&A...635A.139E} for discussion of other metrics). The values of $Z^2$ define the ratio of the systematic error to the statistical error (in a variance sense; $Z=\sqrt{Z^2}$ is the ratio of systematic to statistical error measured by standard deviations). We define $Z^2$ as
\begin{align}
\label{eq: Z-stat}
    Z^2 \equiv \sum_{ l~\rm bins} \Delta C^{\alpha \beta}_{l} {\rm Cov}[C^{\alpha \beta}_{l},C^{\mu \nu}_{l}]^{-1}  \Delta C^{\mu \nu}_{l}
    &= \sum_{l~\rm bins} \frac{N_{modes}}{2}[C^{-1}_{l}]_{\alpha \mu} [C^{-1}_{l}]_{\beta \nu} \Delta C^{\alpha \beta}_{l} \Delta C^{\mu \nu}_{l},
\end{align}
where $N_{modes}$ is the number of $\vec{l}$ modes per $l$-bin, $C^{\alpha \beta}_{l}$ is the shear power spectrum at redshifts $z_\alpha$ and $z_\beta$, ${\rm Cov}[C^{\alpha \beta}_{l},C^{\mu \nu}_{l}]^{-1}$ is the inverse of the covariance matrix between two power spectra functions at different redshift pairs, and the repeated indices implies a summation (except for the $l$ modes). We computed the $Z^2$ statistic of our corrections for the r-band (620 nm) and H-band (1580 nm). These values are listed in the order of their respective wavelengths in Table~\ref{tab: stat_data}. We also computed the ratio of the systematic error ($\Delta C^{\alpha \beta}_{l}$) to the statistical error of the {\slshape Roman} forecast ($\sigma(C^{\alpha \beta}_{l})$) in each bin. The maximum value of this ratio is $0.36$ and $0.044$ for the {\slshape r} and {\slshape H}-band, respectively. This indicates that in $r$-band the systematic error is smaller than the statistical error {\em in each bin}, but when the bins are combined the systematic error is larger as indicated by the $Z^2$ statistic.

We also investigate how the dust systematic impacts measurements of the growth of structure. Using the formalism of Appendix~\ref{sec: Appendix_A}, we split the power spectrum correction $\Delta C_l$ into a bias on the amplitude $\Delta \sigma_8/\sigma_8$ (with all other parameters fixed) and a ``reduced $Z^2$'' analogous to Eq.~(\ref{eq: Z-stat}) but where we only include the residual systematic $\Delta C_l$ that is not degenerate with the change in $\sigma_8$.
Since $Z_{reduced}/Z \sim 0.5$ in both cases, the dust systematic is mostly (but not exactly) degenerate with $\sigma_8$. We can see that the biases are both negative, which makes sense because we expect the effect of dust to underestimate the growth of structure, thus reducing our the value of $\sigma_{8}$.

\begin{table}
\begin{center}
\begin{tabular}{ c|cc| cc| cc} 
 \hline
 Model & \multicolumn2c{$Z$} & \multicolumn2c{$Z_{\rm reduced}$} & \multicolumn2c{$-(\Delta \sigma_{8}/\sigma_8) / 10^{-3}$} \\ [0.5ex] 
 & $H$ band & $r$ band & $H$ band & $r$ band & $H$ band & $r$ band \\
 \hline
 Ceiling & 1.08 & 8.13 & 0.63 & 4.90 & 0.82 & 6.1 \\ 
 \hline
 Base & 0.37 & 2.79 & 0.17 & 1.37 & 0.31 & 2.2 \\
 \hline
 Floor & 0.29 & 2.26 & 0.11 & 0.96 & 0.25 & 1.9 \\ [1ex] 
 \hline
\end{tabular}
\end{center}
\caption{\label{tab: stat_data}The values of $Z$, $Z_{\rm reduced}$ and $\Delta \sigma_{8}/\sigma_{8}$ in the $H$ and $r$ bands.}
\end{table}
% \hline
% Model & $Z^{H,r}$ & $Z^{H,r}_{reduced}$ & $-\Delta \sigma^{H,r}_{8}/\sigma^{H,r}_8 \times 10^{-4}$ \\ [0.5ex] 
% \hline\hline
% Ceiling & (1.08,8.13) & (0.63,4.90) & (0.82,6.1) \\ 
% \hline
% Base & (0.37,2.79) & (0.17,1.37) & (0.37,2.2) \\
% \hline
% Floor & (0.29,2.26) & (0.11,0.96) & (0.25,1.9) \\ [1ex] 
% \hline

%\begin{figure}[h!]
%\caption{The ratio of the corrections with the shear power  spectrum  in the r-band (observer frame) evaluated at the same redshift slice.\label{fig:fractional_r}}
%\end{figure}

%\begin{figure}[h!]
%\plotone{reducedshearRomansdust_r_wt.pdf}
%\caption{Third order corrections to the E-mode shear power spectrum in the r-band (observer frame) of \textit{LSST} (620 nm) evaluated at the same redshift slice.\label{fig:reducedshear_r}}
%\end{figure}

\section{Discussion}
\label{sec:dicussion}

{\slshape Roman}'s most sensitive shape measurement band is the H-band, so in the 2000 deg$^2$ reference survey, our simplest model the dust correction is below the statistical errors. However, if we were to quadruple our observing area to 8000 deg$^2$ (which might happen in an extended mission, or in a single-band extension of the reference survey during the primary mission: \citealt{2021MNRAS.507.1514E}), we would get for $Z_{H} \sim 2\times 0.37=0.74$. As the survey area grows, the need to mitigate the effects of dust will become important. We also see that in the bluer bands, the effects of dust grow. While {\slshape Roman} plans to use the near-infrared filters ($J$ and redder) for shape measurement, the large ground-based surveys have more commonly used $r$ and $i$ filters; this also increases the importance of dust mitigation.

Current cosmic shear results from the Dark Energy Survey Year 3 analysis have reached $\sim2 \%$ precision on $S_8=\sigma_8\sqrt{\Omega_m/0.3}$ (i.e., $\sigma_8$ with a power of $\Omega_m$ scaled out; \citealt{2021arXiv210513543A}).\footnote{The marginalized precision on $\sigma_8$ in \citet{2021arXiv210513543A} is 9\%, but due to the degeneracy direction of $3\times 2$-point constraints, the error on $S_8$ is a better indicator of the constraining power of the current data.} At the present level of precision, we do not expect the dust systematic in our fiducial model to be important. We note that a major current objective is to compare the amplitude of structure measured via weak lensing with that inferred from the cosmic microwave background (CMB) anisotropies. With the primary anisotropy data from \textit{Planck}, $S_8$ is constrained to $\sim 2\%$ and the primordial amplitude $A_s^{1/2}$ is constrained to 0.8\% \citep{2020A&A...641A...6P}, with the latter limited primarily by the uncertainty in the optical depth due to reionization (which rescales the entire high-$l$ CMB power spectrum). So again for the purposes of comparing the primary CMB anisotropies to the weak lensing amplitude, we do not expect  circumgalactic dust to be a significant systematic for the ongoing analyses.

However, looking into the future with {\slshape Roman}, and with the large area optical weak lensing surveys {\slshape Euclid} and LSST, circumgalactic dust will be an important contribution to the data vector ($Z>1$ in the optical and $\sim 0.5$ -- so not negligible -- in the NIR). Moreover, comparisons of low-redshift amplitude of structure to CMB observations will continue to improve (e.g., with CMB-S4; \citealt{2016arXiv161002743A}), including with CMB lensing information where the amplitude is not limited by the optical depth degeneracy.

Fortunately, this same next generation of wide field cosmology surveys will also enable much better constraints on circumgalactic dust. They will provide many more background sources than used in \citet{2010MNRAS.405.1025M}, enabling analyses that are more finely binned by foreground galaxy properties. This includes probing redshift evolution, which could not be constrained in this paper and is one of the main limitations of our study. The combination of these surveys will also have photometric coverage from $u$ band through 2 $\mu$m, which will reduce the sensitivity to the assumed reddening law.

Weak lensing is a powerful statistical tool for understanding the cosmology and dynamics of our Universe. In order to get the most of this powerful tool, we will need to mitigate its systematics, such as extinction of background images due to circumgalatic dust. Our results further motivates the need to better understand circumgalactic dust as well as improve and develop better wide field image surveys such as \textit{Roman} and \textit{LSST}. By better understanding the effects of dust on weak lensing, we will better understand the cosmological models of our Universe.

\section*{Acknowledgements}
Mahalo nui loa to Yi-Kuan Chiang, Jack Elvin-Poole, and Jenna Freudenburg for providing useful discussions regarding analyses. And mahalo to Paulo Montero-Camacho and Bryan Yamashiro for assisting in efficient coding procedures.

This project was supported by the Simons Foundation award 60052667, NASA award 15-WFIRST15-0008, and the David \& Lucile Packard Foundation.

This work made use of the Pitzer supercomputer \citep{Pitzer2018} at the \citet{OhioSupercomputerCenter1987}. 

\appendix

\section{The reduced $Z^{2}$ Metric}
\label{sec: Appendix_A}

Here we describe some details of the error metrics, including the bias on $\sigma_8$.

We start by defining the ``data'' space spanned by the possible power spectra, $\mathbb{D} \equiv \{ C^{\alpha \beta}_{l} \}$ (whose dimenisonality is the number of power spectrum bins). The inverse of the covariance matrix ${\rm Cov}[C_l^{\alpha\beta},C_{l'}^{\mu\nu}]$ defines a metric on this space (the Fisher information metric on ${\mathbb D}$). If there is a systematic $\Delta C_l^{\alpha\beta}$, then $Z^2$ is the square norm of the bias vector in this metric.

\begin{center}
\begin{figure}[h!]
\includegraphics[scale = 0.6]{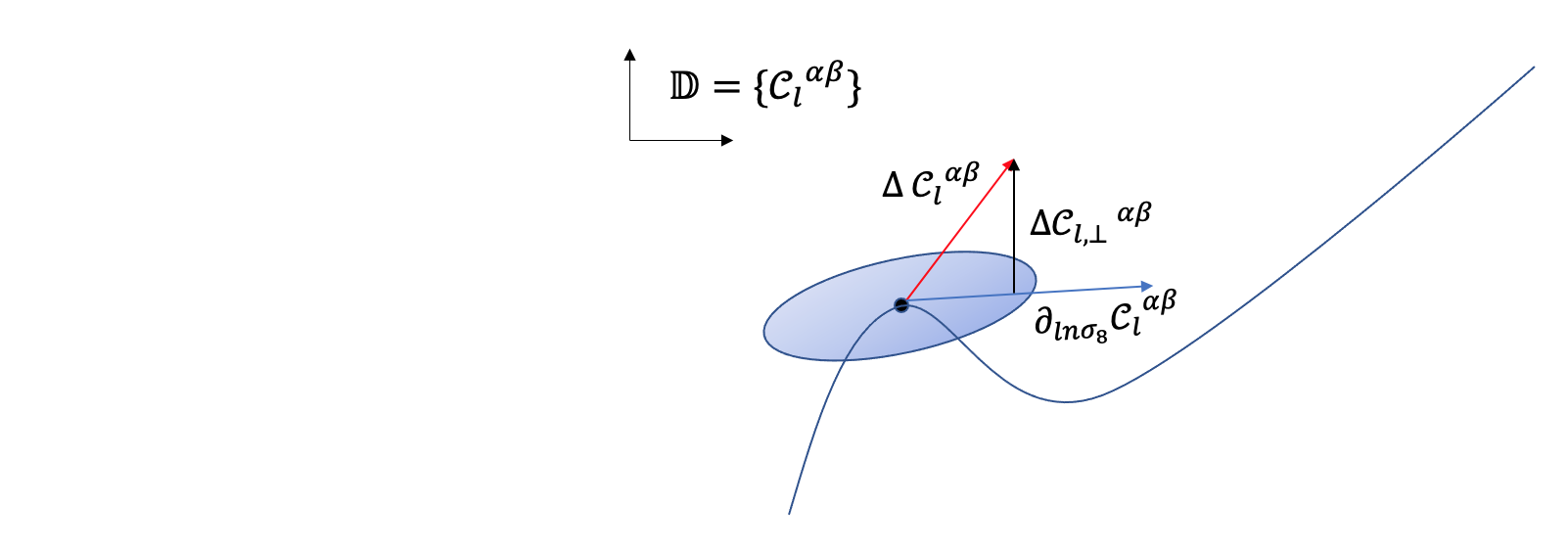}
\caption{A diagrammatic representation relating the different pieces that define the $Z^{2}$ and $Z_{reduced}^2$.\label{fig:data_space}}
\end{figure}
\end{center}

We are now interested in what happens when we fit a single parameter $\ln\sigma_8$. Figure~\ref{fig:data_space} is a two dimensional analog of the higher dimensional space that actually spans our data space. The curve in Fig.~\ref{fig:data_space} is the collection of power spectra with all variables kept fix except that we vary $\ln \sigma_{8}$. If we do a fit varying the single parameter $p = \ln\sigma_8$, then there is a bias of
\begin{align}
\label{eq:delta_epsilon}
    \Delta \ln\sigma_8 = \frac{\sum_{l}\partial_{\ln \sigma_{8}}C^{\alpha \beta}_{l}{\rm Cov}[C^{\alpha \beta}_{l},C^{\mu \nu}_{l}]^{-1} \Delta C^{\mu \nu}_{l}}{\sum_{l}\partial_{\ln \sigma_{8}}C^{\alpha \beta}_{l}{\rm Cov}[C^{\alpha \beta}_{l},C^{\mu \nu}_{l}]^{-1}\partial_{\ln \sigma_{8}}C^{\mu \nu}_{l}}
\end{align}
(see, e.g., Appendix A of \citealt{2004MNRAS.347..909K} or Section 2 of \citealt{2008MNRAS.391..228A}).

In general, the systematic vector $\Delta C_l^{\alpha\beta}$ is not exactly degenerate with $\sigma_8$, i.e., it is not parallel to $\partial C_l^{\alpha\beta}/\partial\ln\sigma_8$ in data space. We are interested in the magnitude of the residual systematic, i.e., $\Delta C_l^{\alpha\beta}$ with the change of $\sigma_8$ scaled out. This leads to:
\begin{gather}
\label{eq:Z_reduced_analytic}
    Z^2_{reduced} = \sum_{l}\left( \Delta C^{\alpha \beta}_{l} - \Delta \epsilon \partial_{\ln \sigma_{8}}C^{\alpha \beta}_{l} \right) {\rm Cov}[C^{\alpha \beta}_{l},C^{\mu \nu}_{l}]^{-1} \left( \Delta C^{\mu \nu}_{l} - \Delta \epsilon \partial_{\ln \sigma_{8}}C^{\mu \nu}_{l} \right),\\
\label{eq:Z_reduced_analytic_1}
    = \sum_{l}\left(\Delta C^{\alpha \beta}_{l} Cov[C^{\alpha \beta}_{l},C^{\mu \nu}_{l}]^{-1} \Delta C^{\mu \nu}_{l} \right) - \frac{(\sum_{l}\partial_{\ln \sigma_{8}}C^{\alpha \beta}_{l}{\rm Cov}[C^{\alpha \beta}_{l},C^{\mu \nu}_{l}]^{-1} \Delta C^{\mu \nu}_{l})^2}{\sum_{l}\partial_{\ln \sigma_{8}}C^{\alpha \beta}_{l}{\rm Cov}[C^{\alpha \beta}_{l},C^{\mu \nu}_{l}]^{-1}\partial_{\ln \sigma_{8}}C^{\mu \nu}_{l}},
\end{gather}
where we see that the first term in Eq. (\ref{eq:Z_reduced_analytic_1}) is the $Z^2$ we defined earlier.

The Fisher information metric provides a simple geometric interpretation of these results. We use the dot product to denote an inner product in this metric. If $\Delta{\bf C}$ is the systematic, and ${\bf v}$ is the vector $\partial C_l^{\alpha\beta}/\partial\ln\sigma_8$, then
\begin{equation}
Z^2 = \Delta{\bf C}\cdot\Delta{\bf C},
~~
\Delta{\bf C}_\perp = \Delta{\bf C} - \frac{{\bf v}\cdot\Delta{\bf C}}{{\bf v}\cdot{\bf v}}{\bf v}
~~{\rm and}~~
Z^2_{\rm reduced} = \Delta{\bf C}_\perp\cdot\Delta{\bf C}_\perp = Z^2\sin^2\theta = Z^2(1-\cos^2\theta),
\end{equation}
where
\begin{equation}
\cos^2\theta = \frac{({\bf v}\cdot\Delta{\bf C})^2}{(\Delta{\bf C}\cdot\Delta{\bf C})({\bf v}\cdot{\bf v})}.
\end{equation}
Thus $Z_{\rm reduced}$ tells us about the part of the systematic orthogonal to changes in $\sigma_8$.

\clearpage
\bibliography{sample63}{}
\bibliographystyle{aasjournal}

\end{document}